\documentstyle[aps,prb]{revtex} 
\begin{document}
\draft
\twocolumn[\hsize\textwidth\columnwidth\hsize\csname
@twocolumnfalse\endcsname

\title{Hydrodynamic approximation for the nonlinear response of a metal surface}
\author{A. Bergara$^1$, J. M. Pitarke$^{1,2}$, and R. H. Ritchie$^3$}
\address{$^1$Materia Kondentsatuaren Fisika Saila, Zientzi Fakultatea,
Euskal Herriko Unibertsitatea,\\ 
644 Posta Kutxatila, 48080 Bilbo, Basque Country, Spain\\
$^2$Donostia International Physics Center (DIPC) and Centro Mixto CSIC-UPV/EHU\\
$^3$Oak Ridge National Laboratory, P.O. Box 2008, Oak Ridge, Tennessee
37831-6123}
\date\today
\maketitle

\begin{abstract}
We present semi-classical and quantized hydrodynamic models to obtain the quadratic
electronic response of a plane-bounded electron gas. Explicit expressions for the
dynamic image potential experienced by charged particles moving
near a jellium surface are derived, up to third order in the projectile charge.
These expressions are employed to compute the image potential at all distances
outside the surface. Though nonlinear corrections are found
to be more important far inside the solid than outside, our results indicate that
the nonlinear image potential is enhanced with respect to the linear image potential by a
factor that is for Al as large as $\sim 1.15$ near the surface in the case of a stationary particle
($v\to 0$) with positive unit charge $e$.
\end{abstract}

] 

\section{Introduction}
\label{sec:level0}

The electronic response of a metal surface to an external perturbation enters a great variety of
important problems in surface science.\cite{Feibelman} Surface spectroscopies employing electrons,
photons, atoms, or ions all involve some kind of electronic excitation at the boundary of the
surface. In particular, the interaction of charged particles with solids has represented an active
field of basic and applied physics,\cite{Echenique} and recently a great amount of research has been
focused on the case of slow ($v<0.5\,v_0$, $v_0$ being the Bohr velocity) highly charged ions
($Z_1>>1$,
$Z_1\,e$ being the ion charge) moving near a solid surface.\cite{Briere,Arnau,High1,High2,High3} For
these projectiles, the parameter
$Z_1v_0/v$ is not small and first-order perturbation or, equivalently, linear-response theories
are not, {\it a priori}, applicable.\cite{Bohr} In the case of charged particles moving inside a
solid, nonlinear effects have been found to be crucial in the interpretation of
energy-loss measurements.\cite{Echenique0,Esbensen,Pitarke0,Pitarke1} Nevertheless, the electronic
response of metal surfaces to the presence of external charged particles, which differs
significantly from that in purely two- or three-dimensional systems, had been described so far
within linear response theory.\cite{Eche}

A central quantity in the interpretation of ion-surface collisions is the so-called image
potential, which represents the interaction between the incoming charge and the polarization
charge that it induces on the surface. In the case of a particle of charge $Z_1e$ located at a
distance $z_0$ far from the surface, into the vacuum, this potential approaches the
long-range classical Coulomb image potential $V^{\rm im}=-Z_1^2e^2/4z_0$.\cite{Jackson} For smaller
values of $z_0$ the image potential differs significantly from its classical limit, the deviation
from the classical result increasing as $z_0$ decreases.\cite{Inkson}

The classical image potential acting between a point classical charge and a metal surface may be
regarded as originating in the coupling with the surface plasmon
field.\cite{Lucas,Ritchie72} Dynamical corrections to the classical image potential have been
discussed in the framework of linear response theory,\cite{Sunjic,Manson,Garcia} and recoil effects
have been treated by including the exchange of virtual excitations between the external charge and
the medium.\cite{Zheng} Preliminary results for nonlinear corrections to the image potential
associated to the quadratic response of solid surfaces have been reported only very
recently.\cite{Javi}   

Theoretical approaches commonly used to describe the electronic
response of jellium surfaces can be classified as being either hydrodynamic\cite{Wilems} in nature
or based on the so-called random-phase approximation (RPA).\cite{Eguiluz} Hydrodynamic approaches
are appealing because of their relative mathematical and computational simplicity, and have been
used with great success in the description of collective phenomena at metal
surfaces.\cite{Ritchie57} Within a hydrodynamic model, one assumes that the collective
motion of the electron gas may be described in terms of the displacement of the electrons from
their original uniform state, and the electron system is characterized by the electron density and
a velocity field. These quantities are then obtained by solving the well-known nonlinear Bloch
hydrodynamic equations. If one does not include quantum properties of the electron system we
refer to the hydrodynamic model as {\it classical}, and as {\it semi-classical} if quantum
properties are introduced through the definition of the internal energy density. If one quantizes
the hydrodynamic hamiltonian on the basis of the existing normal modes,\cite{Barton} we have the
so-called quantized hydrodynamic model which allows to apply standard methods of many-body
perturbation theory.

In a previous work,\cite{Bergara} we used the quantized hydrodynamic model to describe the quadratic
response of a homogeneous electron gas. We derived expressions for the quadratic polarization
induced by a moving charged particle, and demonstrated that they coincide with the plasmon-pole (PP)
approximation\cite{Hedin} to the more accurate quadratic RPA
polarization.\cite{Pitarke2}

In this paper, we first develop semi-classical and quantized hydrodynamic models to
derive the quadratic electronic response of a plane-bounded electron gas, and we then focus on the
evaluation of the nonlinear dynamic image potential experienced by charged particles moving
parallel to a jellium surface. In sections II and III semi-classical and quantized
nonlinear hydrodynamic models are presented, respectively. In section IV numerical calculations of
both linear and quadratic contributions to the image potential are reported, as a function of the
distance from the surface. In section V our conclusions are presented.

\section{Semi-classical Hydrodynamic Model}
\label{sec:level1}

Take an inhomogeneous electron system embedded in a neutralizing ionic background. In the
hydrodynamic limit,\cite{Theory} the total energy of the system can be expressed as\cite{Ying} (we
use atomic units throughout, i. e., $\hbar=m_e=e=1$):
\begin{eqnarray}
H=&&{1\over 2}\int d{\bf r}\,n_e({\bf r},t)\,|{\bf\nabla}\psi({\bf r},t)|^2
-{1\over 2}\int d{\bf r}\,n({\bf r},t)\,V({\bf r},t)\cr\cr
&&-\int d{\bf r}\,n({\bf r},t)\,V_{\rm ext}({\bf r},t)
+\int d{\bf r}\,G\left[n_e({\bf r},t)\right],\label{1.1}
\end{eqnarray}
where irrotational flow has been assumed, i. e., ${\bf u}({\bf r},t)=-{\bf\nabla}\psi({\bf r},t)$,
${\bf u}({\bf r},t)$ being a velocity field, and retardation effects have been neglected. $n_e({\bf
r},t)=n_0({\bf r})+n({\bf r},t)$ is the total electron density, with $n({\bf r},t)$ representing
the deviation from the equilibrium static density $n_0({\bf r})$. $V({\bf r},t)$ is the induced
electric potential, $V_{\rm ext}({\bf r},t)$ represents the external perturbation, and
$G\left[n_e({\bf r},t)\right]$ represents the exchange, correlation and internal kinetic energies
of the electron system. We neglect exchange-correlation contributions to $G\left[n_e({\bf
r},t)\right]$, which we approximate by the Thomas-Fermi functional\cite{Theory}
\begin{equation}
G[n_e({\bf r},t)]={3\over{10}}\,(3\pi^2)^{2/3}\,\left[n_e({\bf r},t)\right]^{5/3}.\label{1.2}
\end{equation}

From Eq. (\ref{1.1}), the basic hydrodynamic equations can be derived, i. e., the continuity
equation,
\begin{equation}
\dot n_e={\bf\nabla}\cdot(n_e\,{\bf\nabla}\psi),\label{1.3}
\end{equation}
and the Bernouilli equation,
\begin{equation}
\eta\,\psi+\dot\psi+\mu={1\over 2}\,|{\bf\nabla}\psi|^2-U+{\delta
G[n_e]\over\delta n_e},\label{1.4}
\end{equation}
which conserves both momentum and energy. Here, $\mu$ is a Lagrangian multiplier (a constant), and
$\eta$ is a positive number representing the internal friction of the electron gas which would
appear as a consequence of the interaction with excitations not included in this description. If the
external perturbation is generated by a charge density $n_{\rm ext}({\bf r},t)$, then the total
electric potential $U({\bf r},t)=V({\bf r},t)+V_{\rm ext}({\bf r},t)$ is obtained from the Poisson
equation
\begin{equation}
\nabla^2U=-4\,\pi\,\left[n_{\rm ext}-n\right].\label{1.5}
\end{equation}

These are nonlinear equations, difficult to solve. Within perturbation theory, we expand the
induced electron density $n({\bf r},t)$ and the velocity potential $\psi({\bf r},t)$ in powers of
the external perturbation,
\begin{equation} 
n=n_1+n_2+...\label{1.6}
\end{equation}
and
\begin{equation}
\psi=\psi_1+\psi_2+...,\label{1.7}
\end{equation}
respectively, assuming that $n_0>>n_1>>n_2>>...$ and $\psi_1>>\psi_2>>...$. Eqs. (\ref{1.3}) and
(\ref{1.4}) are then expanded in powers of the external perturbation, and partial
differential equations for the various orders of $n({\bf r},t)$ and $\psi({\bf r},t)$ are derived.

\subsection{Linear Approximation}
\label{sec:level1A}

Up to first order in the external perturbation one finds, after introduction of Eq. (\ref{1.2})
into Eq. (\ref{1.4}), the linearized hydrodynamic equations
\begin{equation}
\eta\,\psi_1+\dot\psi_1=-U_1+
{\beta^2\over n_0}\,n_1\label{1.A.1}
\end{equation}
and
\begin{equation}
\dot n_1=n_0\,\nabla^2\psi_1,\label{1.A.2}
\end{equation}
where $U_1({\bf r},t)$ is obtained from
\begin{equation}
\nabla^2U_1=-4\,\pi\,\left[n_{\rm ext}-n_1\right],\label{1.A.3}
\end{equation}
and where $\beta=\sqrt{1/3}q_F$, $q_F=(3\pi^2n_0)^{1/3}$ being the Fermi
momentum. Though this is the value of the hydrodynamic speed $\beta$ predicted with use of the
Thomas-Fermi functional $G\left[n_e({\bf r},t)\right]$ of Eq. (\ref{1.2}), the value
$\beta=\sqrt{3/5}q_F$ is expected to be more appropriate when high frequencies of the order of
the plasma frequency are involved.\cite{Ritchie63,Halevi}  

We consider a classical charged particle moving with velocity ${\bf v}$ outside of a semi-infinite
metallic medium, along a trajectory that is parallel to the surface, thereby approximately
simulating the experimental conditions when the projectile approaches the surface at grazing
incidence. Hence, we take the external charge density at ${\bf r}=({\bf r}_\parallel,z)$ to be
given by the following expression:
\begin{equation}
n_{\rm ext}({\bf r},t)=Z_1\,\delta({\bf r}_\parallel-{\bf
v}t)\,\delta(z-z_0),
\end{equation}
the vacuum occupying the half-space $z>0$, and $z_0$ being the distance of the trajectory from the
metal. After Fourier
analyzing both in time and in
${\bf r}_\parallel$, one finds the following linearized hydrodynamic equations with variables
($z;{\bf q},\omega$), ${\bf q}$ being the wave vector parallel to the surface:
\begin{equation}
(-{\rm i}\,\omega+\eta)\,\psi_1=-U_1+{\beta^2\over n_0}\,n_1\label{1.A.4}
\end{equation}
and
\begin{equation}
-{\rm i}\,\omega\,n_1=n_0\,(\psi^{\prime\prime}_1-q^2\,\psi_1),\label{1.A.5}
\end{equation}
where the prime denotes the derivative with respect to $z$, $U_1(z;{\bf q},\omega)$ is obtained
from 
\begin{equation}
U^{\prime\prime}_1-q^2\,U_1=-4\,\pi\,n_{\rm ext}+4\,\pi\,n_1,\label{1.A.6}
\end{equation}
and
\begin{equation}
n_{\rm ext}(z;{\bf q},\omega)=2\,\pi\,Z_1\,\delta(\omega-{\bf q}\cdot{\bf v})\,\delta(z-z_0).
\end{equation}

Assuming a sharp density profile at the surface, i. e., $n_0(z)=n_0\Theta(-z)$ ($\Theta(z)$ is the
Heaviside step function), the appropriate solutions of Eqs. (\ref{1.A.4}) and (\ref{1.A.5}) are:
\begin{equation}
n_1(z;{\bf q},\omega)=n_0\,A_1\,(\Delta^2_{{\bf q},\omega}-q^2)\,{\rm e}^{\Delta_{{\bf
q},\omega}\,z}\,\Theta(-z),\label{1.A.8}
\end{equation}
\begin{equation}
\psi_1(z;{\bf q},\omega)=-{\rm i}\,\omega\,\left(A_1\,{\rm e}^{\Delta_{{\bf
q},\omega}\,z}+B_1\,{\rm e}^{q\,z}\right)\,\Theta(-z),\label{1.A.9}
\end{equation}
\begin{equation}
U_1^<(z;{\bf q},\omega)=\omega_{\rm p}^2\,A_1\,{\rm e}^{\Delta_{{\bf
q},\omega}\,z}+\omega\,(\omega+{\rm i}\eta)\,B_1\,{\rm e}^{q\,z},
\end{equation}
and
\begin{equation}
U_1^>(z;{\bf q},\omega)=C_1\,{\rm e}^{-q\,z}+Z_1\,{4\,\pi^2\over q}\,\delta(\omega-{\bf q}\cdot
{\bf v})\,{\rm e}^{-q\,|z-z_0|}.\label{1.A.7}
\end{equation}
Here, $U_1^{<,>}(z;{\bf q},\omega)$ is the Fourier
transformed electric potential for
$z$ (less than, greater than) zero, $\omega_{\rm p}=({4\pi n_0})^{1/2}$ is the so-called plasma
frequency, and
\begin{equation}
\Delta_{{\bf q},\omega}={1\over\beta}\sqrt{\omega^2_{\rm p}+\beta^2\,q^2-\omega\,(\omega+{\rm
i}\eta)}.
\label{1.A.10}
\end{equation}
The constants $A_1$, $B_1$, and $C_1$ are evaluated from the boundary conditions at the surface:
\begin{equation}
A_1=-{q\over\Delta_{{\bf q},\omega}}\,B_1,
\label{1.A.11}
\end{equation}
\begin{equation}
B_1={8\,\pi^2\,Z_1\over q}\,
{\delta(\omega-{\bf q}\cdot{\bf v})\,{\rm e}^{-q\,z_0}
\over\left[2\,\omega\,(\omega+{\rm i}\,\eta)-\omega_{\rm p}^2\,(1+q/\Delta_{{\bf
q},\omega})\right ]},
\label{1.A.12}
\end{equation}
and
\begin{equation}
C_1={4\,\pi^2\,Z_1\over q}\,
{\omega_{\rm p}^2\,(1-q/\Delta_{{\bf q},\omega})\,\delta(\omega-{\bf q}\cdot{\bf v})\,{\rm
e}^{-q\,z_0}
\over\left[2\,\omega\,(\omega+{\rm i}\,\eta)-\omega_{\rm p}^2\,(1+q/\Delta_{{\bf
q},\omega})\right]}.
\label{1.A.13}
\end{equation}

The induced electric potential $V_1({\bf r},t)$ is the difference between the total and external
potentials $U_1({\bf r},t)$ and $V_{\rm ext}({\bf r},t)$, respectively. The image
potential is defined as half of the induced potential at the position of the projectile
times the projectile charge, and one finds
\begin{eqnarray}
V^{\rm im}_1=&&Z_1^2\,{\omega_{\rm p}^2\over 4\,\pi}\int d^2{\bf q}\,{\rm e}^{-2\,q\,z_0}\cr\cr
&&\times{1/q-1/\Delta_{{\bf q},{\bf q}\cdot{\bf v}}
\over\left[2\,{\bf q}\cdot{\bf v}\,({\bf q}\cdot{\bf v}+{\rm i}\eta)-\omega_{\rm p}^2\,
(1+q/\Delta_{{\bf q},{\bf q}\cdot {\bf v}})\right]}.
\label{1.A.14}
\end{eqnarray}
This agrees with the result obtained using either the specular reflexion (SR)\cite{Ritchiem} or the
semi-classical infinite barrier (SCIB)\cite{Griffin} model of the surface, as long as the
hydrodynamic dielectric response function of the bulk material is used in these models. In
particular, for a stationary charged particle (${\bf v}=0$), one finds
\begin{equation}
V^{\rm im}_1={1\over 2}\,Z_1^2\int dq\,{q-\Delta_{{\bf q},0}\over
q+\Delta_{{\bf q},0}}\,{\rm e}^{-2\,q\,z_0}.
\label{1.A.15}
\end{equation}
This agrees with the early result of Eguiluz.\cite{Eguiluz80}

Within the hydrodynamic model, one can define two characteristic screening lengths $\lambda$. For
a stationary charged particle (${\bf v}=0$), $\lambda=\beta/\omega_{\rm p}$, and in the case of a
swift charged particle ($v>>q_F$), $\lambda=v/\omega_{\rm p}$. In the limit $z_0\to\infty$, the
distance $z_0$ of the projectile from the surface being much larger than the screening length, one
finds the classical image potential 
\begin{equation}
V^{\rm im}_1=-{Z_1^2\over 4\,z_0}. 
\label{1.A.16}
\end{equation}

\subsection{Quadratic Approximation}
\label{sec:level1B}

Up to ${\rm N}$-th order (${\rm N}\ge 2$) in the external perturbation, one finds the hydrodynamic
equations:
\begin{eqnarray}
\eta\psi_{\rm N}+\dot\psi_{\rm N}=&&
{1\over 2}\sum^{{\rm N}-1}_{{\rm m}=1}\left({\bf \nabla}\psi_{{\rm N}-{\rm m}}\cdot{\bf
\nabla}\psi_{{\rm m}}  
\right)-V_{\rm N}\cr\cr
&&+{3\over 2}\,\beta^2\,n_0^{-2/3}\sum^{\rm N}_{{\rm k}=1}
\left[{\prod^{\rm k}_{{\rm s}=1}(5-3{\rm s})\over n_0^{{\rm k}-2/3}3^{\rm k}}\right.\cr\cr
&&\times\left.\sum^{\rm N}_{{{{\rm k}_1,{\rm k}_2,...,{\rm k}_{\rm N}=0}\atop {\sum^{\rm
N}_{{\rm i}=1}{\rm k}_{\rm i}={\rm k} ;\
\sum^{\rm N}_{{\rm i}=1}{\rm ik_i=N}}} }
\left(\prod^{\rm N}_{{\rm j}=1}{n_{\rm j}^{{\rm k_j}}\over{\rm k_j}!}\right)\right] 
\label{1.B.1}
\end{eqnarray}
and
\begin{equation}
\dot n_{\rm N}=n_0\nabla^2\psi_{\rm N}+\sum^{{\rm N}-1}_{{\rm m}=1} 
\left[{n_{{\rm N-m}}\nabla^2\psi_{\rm m}+{\bf\nabla}n_{{\rm N-m}}\cdot{\bf\nabla}\psi_{{\rm m}}}
\right],
\label{1.B.2}
\end{equation}
where $V_{\rm N}({\bf r},t)$ is obtained from
\begin{equation}
\nabla^2V_{\rm N}=4\,\pi\,n_{\rm N}.
\label{1.B.3}
\end{equation}

After Fourier transforming in time and in ${\bf r}_\parallel$, we obtain for ${\rm N}=2$ the
quadratic approximation
\begin{equation}
-{\rm i}\,(\omega+{\rm i}\,\eta)\,\psi_2=-V_2+ 
{1\over 2}\,\left|{\bf\nabla}\psi_1\right|^2_{(z;{\bf q},\omega)}+
{\beta^2\over n_0}\,n_2-{\beta^2\over 6\,n^2_0}\,n_1^2
\label{1.B.4}
\end{equation}
and
\begin{equation}
-{\rm i}\,\omega\,n_2=n_0(\psi^{\prime\prime}_2-q^2\psi_2)+
\left[n_1\nabla^2\psi_1+{\bf\nabla}n_1\cdot{\bf\nabla}\psi_1\right]_{(z;{\bf q},\omega)},
\label{1.B.5}
\end{equation}
where the prime denotes the derivative with respect to $z$, and the Fourier transformed
potential $V_2(z;{\bf q},\omega)$ is obtained from 
\begin{equation}
V^{\prime \prime}_2-q^2\,V_2=4\,\pi\,n_2.\label{1.B.6}
\end{equation}

In particular, for a stationary (${\bf v}=0$) charged particle Eqs. (\ref{1.B.4}) and
(\ref{1.B.5}) yield 
\begin{equation}
 V_2 = {{\beta^2}\over{n_0}} n_2 - {{\beta^2}\over{6 n^2_0}} n_1^2,\label{1.B.7}
\end{equation}
where $V_2(z;{\bf q},\omega)$ is still obtained from Eq. (\ref{1.B.6}). Solving these equations
with the assumption of a sharp density profile at the surface, quadratic contributions to the
induced density and potential are found to be given by the following expressions:
\begin{eqnarray}
n_2(z;{\bf q},\omega)&&=\Theta(-z)\,\left[A_2\,{\rm e}^{\Delta_{{\bf q},0}\,z}\right.\cr\cr
&&+Z_1^2\,{1\over 6\,n_0}\int{d^2{\bf k}\over(2\pi)^2}\,{\rm e}^{\left(\Delta_{{\bf
k},0}+\Delta_{{\bf k}-{\bf q},0}\right)\,z}\cr\cr
&&\left.\times{\left(\Delta_{{\bf k},0}+
\Delta_{{\bf k}-{\bf q},0}\right)^2-q^2\over\left(\Delta_{{\bf k},0}+
\Delta_{{\bf k}-{\bf q},0}\right)^2-\Delta^2_{{\bf q},0}}\,f_{\bf k}\,f_{{\bf k}-{\bf
q}}\right],
\label{1.B.12}
\end{eqnarray}
\begin{eqnarray}
V_2^<&&(z;{\bf q},\omega)={4\,\pi\over\Delta^2_{{\bf q},0}-q^2}\,A_2\,{\rm e}^{\Delta_{{\bf
q},\omega}\,z}+B_2\,{\rm e}^{q\,z}\cr\cr
&&+Z_1^2\,{4\,\pi\over 6n_0}\int{d^2{\bf k}\over(2\pi)^2}
{f_{\bf k}\,f_{{\bf k}-{\bf q}}\,{\rm e}^{\left(\Delta_{{\bf k},0}+\Delta_{{\bf k}-{\bf
q},0}\right)\,z}\over\left(\Delta_{{\bf k},0}+
\Delta_{{\bf k}-{\bf q},0}\right)^2-\Delta^2_{{\bf q},0}},
\label{1.B.9}
\end{eqnarray}
and
\begin{equation}
V_2^>(z;{\bf q},\omega)=C_2\,{\rm e}^{q\,z},
\end{equation}
where
\begin{equation}
f_{\bf q}=\left(\Delta_{{\bf q},0}-q\right)\,{\rm e}^{-q\,z_0},
\label{1.B.11}
\end{equation}
and $\Delta_{{\bf q},\omega}$ is obtained from Eq. (\ref{1.A.10}). The constants $A_2$, $B_2$ and
$C_2$ are evaluated from the boundary conditions at the surface.

Hence, we find the quadratic contribution to the image potential of a stationary (${\bf v}=0$)
charged particle to be given by the following expression:
\begin{eqnarray}
V^{\rm im}_2=&&-Z_1^3{\beta^2\over{6\,\pi\,\omega_{\rm p}^4}}\int^{\infty}_0dq\,q\int^{\infty}_0
dq_1\,q_1\cr\cr
&&\times\int^{2\pi}_0d\theta\,{f_{\bf q}\,f_{{\bf q}_1}\,f_{{\bf
q}-{{\bf q}_1}}\over\left(\Delta_{{\bf q},0}+\Delta_{{{\bf q}_1},0}+\Delta_{{{\bf q}-
{{\bf q}_1}},0}\right)},
\label{1.B.15}
\end{eqnarray}
$\theta$ being the angle between ${\bf q}$ and ${{\bf q}_1}$. In the case of a non-dispersive
electron gas ($\beta=0$), this contribution to the image potential vanishes. On the other hand, in
the limit as $z_0\to\infty$ only the low-momentum form of the integrand of Eq. (\ref{1.B.15})
contributes to the integration, and we find in this limit
\begin{eqnarray}
V^{\rm im}_2=&&-Z^3_1\,{1\over 18\,\pi\omega_{\rm p}^2}\int^{\infty}_0 dq\,q
\int^{\infty}_0dq_1\,q_1\,{\rm e}^{-\left(q+q_1\right)\,z_0}\cr\cr
&&\times\int^{2\pi}_0d\theta\,{\rm e}^{-\left|{\bf q}-{{\bf q}_1}\right|\,z_0}.
\label{1.B.16}
\end{eqnarray}
Numerical integration yields
\begin{equation}
V^{\rm im}_2=-Z_1^3{0.82\over 18\,\pi\omega_{\rm p}^2\,z_0^4},
\label{1.B.17}
\end{equation}
and, therefore (see Eq. (\ref{1.A.16})),
\begin{equation}
V^{\rm im}=-{Z_1^2\over 4\,z_0}\,\left[1+Z_1\,{1.93\times
10^{-2}\over(z_0/r_s)^3}+O\left(Z_1^2\right)\right].
\label{1.B.18}
\end{equation}
This is a reasonable approximation for the image potential of a stationary charge, as long as the
distance from the surface $z_0$ is larger than the characteristic screening length
$\beta/\omega_{\rm p}$, i. e., for $z_0{>\atop\sim}\sqrt{r_s\,a_0}$, $a_0=\hbar^2/(m_e\,e^2)$ being
the Bohr radius. We note that in the high-density limit ($r_s\to 0$) there is no quadratic
contribution to the image potential of a stationary charge, while at metallic densities ($r_s\sim
2-6$) quadratic corrections might give rise to an image potential that is at
$z_0\sim\sqrt{r_s\,a_0}$ larger than the linear image potential by a factor as large as
$1.05-1.3$ in the case of a stationary particle with positive unit charge $e$ ($Z_1=1$).

\section{Quantized Hydrodynamic Model}
\label{sec:level2}

Within a quantized hydrodynamic model of the electron gas, we first expand the hamiltonian of
Eq. (\ref{1.1}) in powers of the induced electron density $n({\bf r},t)$. After introduction of
the Thomas-Fermi functional of Eq. (\ref{1.2}) into Eq. (\ref{1.1}), up to third order one finds
\begin{equation}
 H=H_G+H_0+H_1+H_{\rm ext}\label{2.1},
\end{equation}
where
\begin{equation}
H_G={3\over10}\,\left(3\pi^2\right)^{2/3}\,n_0^{5/3},
\label{2.2}
\end{equation}
\begin{equation}
H_0=\int d{\bf
r}\left[{1\over2}\,n_0\,|{\bf\nabla}\psi|^2+{\beta^2\over2\,n_0}\,n^2-{1\over2}\,n\,V\right],
\label{2.3}
\end{equation}
\begin{equation}
H_1=\int d{\bf r}\left[{1\over2}\,n\,|{\bf \nabla}\psi|^2-
{\beta^2\over18\,n^2_0}\,n^3\right],\label{2.4}
\end{equation}
and 
\begin{equation}
H_{\rm ext}=-\int d{\bf r}\,n\,V_{\rm ext}.\label{2.5}
\end{equation}
$H_G$ is the Thomas-Fermi ground state of the static unperturbed electron system,
$H_0$ represents the linear deviation from the ground state, $H_1$ appears as a consequence of the
nonlinearity of the electron system, and $H_{\rm ext}$ represents the contribution to the
hamiltonian from the coupling with the external charged particle.

We consider, as in the previous section, a semi-infinite electron system embedded in a
neutralizing ionic background, assuming a sharp electron-density profile at the surface. For each
value of ${\bf q}$ (the wave vector parallel to the surface) there exist both bulk and surface
normal modes of oscillation with frequencies given by the following dispersion
relations:
\begin{equation}
(\omega^B_{{\bf q},p})^2=\omega_{\rm p}^2+\beta^2\,(q^2+p^2)\label{2.6}
\end{equation}
and
\begin{equation}
(\omega^S_{\bf q})^2={1\over 2}\left[\omega_{\rm p}^2+\beta^2\,q^2+\beta\,q\,(2\,\omega_{\rm p}^2
+\beta^2\,q^2)^{1/2}\right],\label{2.7}
\end{equation}
respectively, where $\beta$ represents the speed of propagation of
hydrodynamic disturbances in the electron system. As in the previous section, we
choose $\beta=\sqrt{3/5}q_F$.

Now we follow Ref.\onlinecite{Barton} to quantize the hamiltonian of Eq. (\ref{2.1}) on the basis
of the normal modes corresponding to Eqs. (\ref{2.6}) and (\ref{2.7}), which we shall refer after
quantization as bulk and surface plasmons, respectively. We find
\begin{equation}
H_0=H_0^S+H_0^B,\label{2.8}
\end{equation}
$H_0^B$ and $H_0^S$ being free bulk and surface plasmon hamiltonians, respectively:
\begin{equation}
H^B_0={1\over\Omega}\sum_{{\bf q},p>0}[1/2+\omega^B_{{\bf q},p}]\
a_{{\bf q},p}^{\dag}(t)\ a_{{\bf q},p}(t)\label{2.9}
\end{equation}
and
\begin{equation}
H^S_0={1\over A}\sum_{\bf q}\left[1/2+\omega^S_{\bf q}\right]\
b_{{\bf q}}^{\dag}(t)\  
b_{{\bf q}}(t).\label{2.10}
\end{equation}
Here $\Omega$ and $A$ represent the normalization volume and the normalization
area of the surface, respectively, and $a_{{\bf q},p}(t)$ and $b_{{\bf q}}(t)$ are
Bose-Einstein operators that annihilate bulk and surface plasmons with wave vectors $({\bf
q},p)$ and ${\bf q}$, respectively. The quantized $H_1$ hamiltonian, which contains the quadratic
electronic response of the electron system, will be consider below. For the hamiltonian containing
the coupling between the external particle and either bulk or surface plasmon fields, one finds
\begin{equation}
H_{\rm ext}=H_{\rm ext}^S+H_{\rm ext}^B,\label{2.11}
\end{equation}
where
\begin{equation} 
H_{\rm ext}^{B/S}=\int d{\bf r}\,\rho_{\rm ext}({\bf r},t)\,\phi^{B/S}({\bf r},t),\label{2.12}  
\end{equation}
$\phi^{B/S}({\bf r},t)$ representing operators corresponding to the scalar electric potential
due to bulk/surface plasmons. Outside the metal ($z>0$),  
\begin{equation}
\phi^B({\bf r},t)=-{1\over\Omega}\sum_{{\bf q},p>0}f_{{\bf q},p}^B(z)\, 
{\rm e}^{{\rm i}{\bf q}\cdot{\bf r}_\parallel}
\,\chi^B_{{\bf q},p}(t)\label{2.13}
\end{equation}
and
\begin{equation}
\phi^S({\bf r},t)=-{1\over A}\sum_{\bf q}f_{\bf q}^S(z)
\,{\rm e}^{{\rm i}{\bf q}\cdot{\bf r}_\parallel}
\,\chi^S_{{\bf q}}(t),\label{2.14}
\end{equation}
$\chi^B_{{\bf q},p}(t)$ and $\chi^S_{{\bf q}}(t)$ representing operators associated to the electron
density induced by the excitation of bulk and surface plasmon fields, respectively,
\begin{equation}
\chi^B_{{\bf q},p}(t)=a_{{\bf q},p}^{\dag}(t)+a_{-{\bf q},p}(t)\label{2.14.A}
\end{equation}
and 
\begin{equation}
\chi^S_{{\bf q}}(t)=b_{{\bf q}}^{\dag}(t)+b_{-{\bf q}}(t),\label{2.14.B}
\end{equation}
and $f_{{\bf q},p}^B(z)$ and $f_{\bf q}^S(z)$ being bulk and surface coupling functions,
\begin{equation}
f_{{\bf q},p}^B(z)={\sqrt{2\pi/\omega^B_{{\bf q},p}}\,\omega_{\rm p}\,p\,{\rm
e}^{-q\,z}\over
\left[p^4+p^2(q^2+\omega_{\rm p}^2/\beta^2)+\omega_{\rm
p}^4/(4\beta^4)\right]^{1/2}}\label{2.15}
\end{equation}
and 
\begin{equation}
f_{\bf q}^S(z)={\sqrt{\pi\gamma_{\bf q}/\omega^S_{\bf q}}\,\omega_{\rm p}\over
\left[q\,\left(q+2\gamma_{\bf q}\right)\right]^{1/2}}\,{\rm e}^{-q\,z}.\label{2.16}
\end{equation}
Here $\gamma_{\bf q}$ represents the so-called inverse decay length of surface plasmon charge
fluctuations,\cite{Eguiluz80}
\begin{equation}
\gamma_{\bf q}={1\over 2\beta}\left[-\beta
q+\sqrt{2\omega_{\rm p}^2+\beta^2q^2}\right].\label{2.17}
\end{equation}
In the absence of electron-gas dispersion ($\beta=0$), the scalar electric potential
$\phi^B({\bf r},t)$ due to bulk plasmons vanishes outside the surface; hence, in this case probes
exterior to the solid can only generate surface excitations.

We derive the potential induced by the presence of the external perturbing
charge as the expectation value of the total scalar potential
operator,\cite{Fetter}
\begin{equation}
V({\bf r},t)={<\Psi_0|\phi_H^B+\phi_H^S|\Psi_0>\over<\Psi_0|\Psi_0>},\label{2.18}
\end{equation}
where $|\Psi_0>$ is the Heisenberg ground state of the interacting system, and
where $\phi_H^B({\bf r},t)$ and $\phi_H^S({\bf r},t)$ are the operators of
Eqs. (\ref{2.13}) and (\ref{2.14}) in the Heisenberg picture. Eq. (\ref{2.18}) can be rewritten as
follows
\begin{equation}
V({\bf r},t)={{<\Phi_0|U^{\dag}
(t,-\infty)[\phi_I^B+\phi_I^S]U(t,-\infty)|\Phi_0>}\over{<\Phi_0|U^{\dag}
(t,-\infty)U(t,-\infty)|\Phi_0>}},\label{2.19}
\end{equation}
where $|\Phi_0>$ represents the ground state of an interacting electron system described by
the free plasmon hamiltonian $H_0$, and $U(t_1,t_0)$ is the evolution operator,
\begin{equation}
U(t_1,t_0)=T\left\{{\rm exp}\left[-{\rm i}\int_{t_0}^{t_1}dt\,\left(H^I_1+H^I_{\rm
ext}\right)\right]\right\},\label{2.20}
\end{equation}
$T$ being the chronological operator, and $H^I_1$/$H^I_{\rm ext}$ representing the
hamiltonians $H_1$/$H_{\rm ext}$ in the interaction picture.

\subsection{Linear Approximation}
\label{sec:level2A}
Up to first order in the external perturbation one finds, after introduction of Eq. (\ref{2.20})
into Eq. (\ref{2.19}), the linear contribution to the induced potential:
\begin{equation}
V_1({\bf r},t)=V_1^B+V_1^S,\label{2.A.1}
\end{equation}
where
\begin{eqnarray}
V^B_1({\bf r},t)=&&{Z_1}\int{d^2{\bf q}\over(2\pi)^2}
\int_0^{\infty}{dp\over2\pi}\,f^B_{{\bf q},p}(z)\,f^B_{{\bf q},p}(z_0)\cr\cr &&\times
D^B_{{\bf q},p}({\bf q}\cdot{\bf v})\,{\rm e}^{{\rm i}\,{\bf q}\cdot({\bf r}_\parallel-{\bf
v}\,t)}\label{2.A.2}
\end{eqnarray}
and
\begin{eqnarray}
V^S_1({\bf r},t)=&&{Z_1}\int{d^2{\bf
q}\over(2\pi)^2}\,f^S_{\bf q}(z)\,f^S_{\bf q}(z_0)\cr\cr
&&\times D^S_{\bf q}({\bf q}\cdot{\bf
v})\,{\rm e}^{{\rm i}\,{\bf q}\cdot({\bf r}_\parallel-{\bf v},t)},\label{2.A.3}
\end{eqnarray}
$D^B_{{\bf q},p}(\omega)$ and $D^S_{\bf q}(\omega)$ representing retarded Green's functions for
the operators $\chi^B_{{\bf q},p}(t)$ and $\chi^S_{\bf q}(t)$, respectively:
\begin{equation}
D^B_{{\bf q},p}(\omega)={2\,\omega^B_{{\bf q},p}\over\omega\,(\omega+
{\rm i}\,\eta)-(\omega^B_{{\bf q},p})^2}\label{2.A.4}
\end{equation}
and 
\begin{equation}
D^S_{\bf q}(\omega)={2\,\omega^S_{\bf q}\over\omega\,(\omega+
{\rm i}\,\eta)-(\omega^S_{\bf q})^2}.\label{2.A.5}
\end{equation}
Eq. (\ref{2.A.1}) agrees with the linear contribution to the induced potential obtained, within
the semi-classical hydrodynamic model, as the difference between the Fourier transform of the total
potential of Eq. (\ref{1.A.7}) and the external potential $V_{\rm ext}({\bf r},t)$. Within the
semi-classical approach, the role played by bulk and surface plasmons goes unnoticed; however, the
quantized hydrodynamic model provides explicit separate expressions for the contributions to the
induced potential coming from the coupling with bulk and surface plasmons. The role that bulk and
surface plasmon excitation plays on the energy loss of charged particles interacting with metal
surfaces has been investigated recently,\cite{New} showing that bulk plasmons are excited even in
the case of charged particles that do not penetrate into the solid.

\subsection{Quadratic Approximation}
\label{sec:level2B}
Quantizing the hamiltonian $H_1$ of Eq. (\ref{2.4}) on the basis of both bulk and surface plasmons
is a difficult task, because of the interaction between bulk and surface plasmon fields. Hence,
for the description of the nonlinear response to a charged particle moving outside of a
semi-infinite medium, now we neglect bulk-plasmon contributions and find the following expression
for the quantized $H_1$ hamiltonian:
\begin{eqnarray}
H_1={1\over A^2}&&\sum_{{\bf q}_1}\sum_{{\bf q}_2}
\left[-\Lambda_{{\bf q}_1,{\bf q}_2}\,\dot\chi^S_{{\bf q}_1}(t)\,\dot\chi^S_{{\bf
q}_2}(t)\right.\cr\cr
&&\left.+{\cal P}_{{\bf q}_1,{\bf q}_2}\,\chi^S_{{\bf q}_1}(t)\,\chi^S_{{\bf
q}_2}(t)\right]\chi^S_{-({\bf q}_1+{\bf q}_2)}(t),\label{2.B.1}
\end{eqnarray}
where
\begin{eqnarray}
&&\Lambda_{{\bf q}_1,{\bf q}_2}={\gamma_{{\bf q}_1+{\bf q}_2}+|{\bf q}_1+{\bf
q}_2|\over\left(\gamma_{{\bf q}_1}-q_1\right)\left(\gamma_{{\bf q}_2}-q_2\right)\sqrt{2\,n_0
\,q_1\,q_2\,\omega^S_{{\bf q}_1}\,\omega^S_{{\bf q}_2}\,\omega^S_{{\bf q}_1+{\bf q}_2}}}\cr\cr\cr
&&\times\left[|{\bf q}_1+{\bf q}_2|\,\gamma_{{\bf q}_1+{\bf
q}_2}\,\gamma_{{\bf q}_1}\,\gamma_{{\bf q}_2}\over\left(q_1+2\gamma_{{\bf q}_1}\right)\,
\left(q_2+2\gamma_{{\bf q}_2}\right)\,\left(|{\bf q}_1+{\bf q}_2|+2\gamma_{{\bf q}_1+{\bf
q}_2}\right)\right]^{1/2}\cr\cr\cr
&&\times\left[{q_1\,q_2\,\left({\bf q}_1\cdot{\bf q}_2-\gamma_{{\bf q}_1}\gamma_{{\bf
q}_2}\right)\over
\gamma_{{\bf q}_1}+\gamma_{{\bf q}_2}+\gamma_{{\bf q}_1+{\bf q}_2}}
-{q_1\,\gamma_{{\bf q}_2}\,\left({\bf q}_1\cdot{\bf q}_2-\gamma_{{\bf q}_1}\,q_2\right)\over
\gamma_{{\bf q}_1}+q_2+\gamma_{{\bf q}_1+{\bf q}_2}}\right.\cr\cr\cr
&&\left.+{q_2\,\gamma_{{\bf q}_1}\left({\bf q}_1\cdot{\bf q}_2-q_1\,\gamma_{{\bf q}_2}\right)\over
q_1 +\gamma_{{\bf q}_2}+\gamma_{{\bf q}_1+{\bf q}_2}}-{\gamma_{{\bf q}_1}\gamma_{{\bf
q}_2}\left({\bf q}_1\cdot{\bf q}_2-q_1\,q_2\right)\over q_1+q_2+\gamma_{{\bf q}_1+{\bf
q}_2}}\right]\label{2.B.2}
\end{eqnarray}
and
\begin{eqnarray}
&&{\cal P}_{{\bf q}_1,{\bf q}_2}=
-{{\beta^2}\over{18 n^{1/2}_0}}
\left[{q_1\,q_2\,|{\bf q}_1+{\bf q}_2|\,\gamma_{{\bf q}_1}\over\omega^S_{{\bf
q}_1}\,\omega^S_{{\bf q}_2}
\,\omega^S_{{\bf q}_1+{\bf q}_2}\,\left(q_1+2\,\gamma_{{\bf q}_1}\right)}\right.\cr\cr
&&\times\left.{\gamma_{{\bf q}_2}\,\gamma_{{\bf q}_1+{\bf q}_2}\over\left(q_2+2\gamma_{{\bf
q}_2}\right)
\left(|{\bf q}_1+{\bf q}_2|+2\,\gamma_{{\bf q}_1+{\bf q}_2}\right)}\right]^{1/2}\cr\cr\cr
&&\times{\left(q_1+\gamma_{{\bf q}_1}\right)\left(q_2+\gamma_{{\bf q}_2}\right)\left(|{\bf q}_1+{\bf
q}_2|+\gamma_{{\bf q}_1+{\bf q}_2}\right)\over\gamma_{{\bf q}_1}+\gamma_{{\bf q}_2}+\gamma_{{\bf
q}_1+{\bf q}_2}}.\label{2.B.3}
\end{eqnarray}
The first term in Eq. (\ref{2.B.1}) comes from the kinetic energy of fluid flow, $\int d{\bf
r}\,n\,|{\bf\nabla}\psi|^2/2$, while the second term, which is proportional
to $\beta^2$, comes from the internal energy $G[n]$ of Eq. (\ref{1.2}).

Introducing Eqs. (\ref{2.11}) and (\ref{2.B.1}) into Eq. (\ref{2.20}), and Eq. (\ref{2.20}) into Eq.
(\ref{2.19}), the quadratic contribution to the induced potential is found to be given by the
following expression:
\begin{eqnarray}
V_2&&({\bf r},t)=-Z_1^2
\int{d^2{\bf q}\over(2\pi)^2}{d^2{\bf q}_1\over(2\pi)^2}\,f^S_{\bf q}(z+z_0)\,f^S_{{\bf
q}_1}(z_0)\cr\cr &&\times f^S_{{\bf q}-{\bf q}_1}(z_0)\,D^S_{\bf q}(\omega)\,D^S_{{\bf
q}_1}(\omega_1)\,D^S_{{\bf q}-{\bf q}_1}(\omega-\omega_1)\cr\cr
&&\times\left[\omega\,\omega_1\,\Lambda_{-{\bf q},{\bf
q}_1}-\omega\,(\omega-\omega_1)\,\Lambda_{-{\bf q},{\bf q}-{\bf q}_1}\right.\cr\cr
&&+\left.\omega_1\,(\omega-\omega_1)\,\Lambda_{{\bf q}_1,{\bf q}-{\bf q}_1}
+3\,{\cal P}_{{\bf q},-{\bf q}_1}\right]
\,{\rm e}^{{\rm i}\,{\bf q}\cdot({\bf r}_\parallel-{\bf v} t)},\label{2.B.4}
\end{eqnarray}
where $\omega={\bf q}\cdot{\bf v}$ and $\omega_1={\bf q}_1\cdot{\bf v}$. Linear and quadratic
contributions to the induced potential (see Eqs. (\ref{2.A.1}) and (\ref{2.B.4})) can be
represented diagrammatically as in Fig. 1. As for the quadratic contribution, the external
perturbation (white circles) acts twice on the electron gas through plasmon propagators (wavy
lines), thereby creating an induced potential at point ${\bf r}$ and time $t$ (crosses).

In particular, for a stationary charged particle we have $\omega=\omega_1=0$, thereby only the
internal kinetic energy of Eq. (\ref{1.2}) contributing to the quadratic induced potential. For the
quadratic contribution to the image potential of a stationary charged particle we find
\begin{eqnarray}
V^{\rm im}_2=&&-Z_1^3{\beta^2\,\omega_{\rm p}^2\over 6\,\pi}\int_0^\infty
dq\,q\int_0^\infty dq_1\,q_1\,{\rm e}^{-(q+q_1)\,z_0}\cr\cr\cr
&&\times\int^{2\pi}_0d\theta\,{\gamma_{\bf q}\,\gamma_{{\bf q}_1}\,\gamma_{{\bf q}-{\bf q}_1}\,
{\rm e}^{-\left({\bf q}-{\bf q}_1\right)\,z_0}
\over\left(\omega^S_{\bf q}\,\omega^S_{{\bf q}_1}\,\omega^S_{{\bf q}-{\bf q}_1}\right)^2 
\,\left(\gamma_{\bf q}+\gamma_{{\bf q}_1}+\gamma_{{\bf q}-{\bf q}_1}\right)}\cr\cr\cr
&&\times{\left(q+\gamma_{\bf q}\right)\,\left({\bf q}_1+\gamma_{{\bf q}_1}\right)\,\left(|{\bf
q}-{\bf q}_1|+\gamma_{{\bf q}-{\bf q}_1}\right)
\over\left(q+2\gamma_{\bf q}\right)\,\left(q_1+2\gamma_{{\bf q}_1}\right)\,    
\left(|{\bf q}-{\bf q}_1|+2\gamma_{|{\bf q}-{\bf q}_1}\right)},
\label{2.B.5}
\end{eqnarray}
which in the case of a non-dispersive electron gas ($\beta=0$) vanishes. Here $\theta$ is the angle
between ${\bf q}$ and ${\bf q}_1$, and in the limit as $z_0\to\infty$ Eq. (\ref{2.B.5}) yields
\begin{equation}
V^{\rm im}_2=-Z_1^3{0.41\over 18\,\pi\omega_{\rm p}^2\,z_0^4}.
\label{2.B.6}
\end{equation}
This quadratic contribution to the image potential is half the result obtained within the
semi-classical hydrodynamic model (see Eq. (\ref{1.B.17})). In the case of a stationary charged
particle the whole quadratic contribution to the image potential comes from the second term in
Eq.(\ref{2.4}), i. e., from the linearly induced electron density acting twice on the external
charge. The total electron density $n_1$ induced at the surface by a stationary charged particle
is, in the limit $z_0\to\infty$, $\sqrt{2}$ times the electron density $n_1^S$ induced through
coupling with surface plasmons (see Appendix A). As a consequence, the total quadratic
contribution to the image potential is in this limit (see Eq. (\ref{1.B.17})) twice as large as the
quadratic contribution of Eq. (\ref{2.B.6}), which has been deduced by neglecting the coupling
with bulk plasmons.

\section{Results}
\label{sec:level3}
Fig. 2 shows plots of the linear contribution to the image potential of a particle with unit
charge $e$ ($Z_1=1$) traveling parallel to the
surface of a semi-infinite electron gas characterized by a static electron density $n_0$ equal to
the average electron density in the conduction band of aluminum ($r_s\sim 2$).\cite{note1} These
plots are shown as a function of $z_0$, the distance from the surface, and the speed is
taken to be $v=0$ (Fig. 2a) and $v=2$ (Fig. 2b). Contributions from couplings with bulk and
surface plasmon fields, as obtained from Eqs. (\ref{2.A.2}) and (\ref{2.A.3}), respectively, are
represented separately by dashed and dotted lines, and the total linear contribution
to the image potential, obtained from either Eq. (\ref{1.A.14}) or Eq. (\ref{2.A.1}), is represented
by a solid line. For comparison, the classical image potential of Eq. (\ref{1.A.16}) is represented
by a dashed-dotted line, showing that it converges with the full linear result when the distance
$z_0$ is well above the screening length, i. e., $z_0{>\atop\sim}\sqrt{r_s\,a_0}$ for $v=0$ and
$z_0{>\atop\sim}2\,\sqrt{r_s\,a_0}$ for $v=2$. 

Quadratic contributions to the image potential of a particle with unit charge $e$ ($Z_1=1$)
traveling, as in Fig. 2, parallel to the surface of a plane-bounded electron gas are
depicted in Fig. 3. The electron-density parameter $r_s$ and the velocity of the external charged
particle are the same as those considered in Fig. 2. In the case of a stationary charged particle
(Fig. 3a), the contribution from coupling with the surface plasmon field, as obtained from Eq.
(\ref{2.B.5}), is represented (dotted line) together with the total quadratic contribution (solid
line), obtained from Eq. (\ref{1.B.15}). At large distances from the surface ($z_0\to\infty$), the
total quadratic contribution to the image potential is twice as large as the quadratic contribution
from surface plasmon excitation, as discussed after Eq. (\ref{2.B.6}). At smaller values of
$z_0$, the quadratic contribution from the bulk channel becomes dominant, the total quadratic
image potential being near the surface larger than the quadratic contribution from the surface
channel by a factor of $\sim 10$. The
quadratic contribution from the surface channel to the image potential of a charged particle moving
with speed $v=2$, as obtained from Eq. (\ref{2.B.4}), is represented by a dotted line in Fig. 3b.

Fig. 4 exhibits by a solid line, as a function of the distance $z_0$ from the surface, the ratio
between full quadratic (solid line of Fig. 3a) and linear (solid line of Fig. 2a) contributions to
the image potential of a stationary particle with unit charge $e$ ($Z_1=1$). For comparison, the
ratio $Z_1\,1.93\times10^{-2}\,(z_0/r_s)^{-3}$, as obtained from Eq. (\ref{1.B.18}), is represented
by a dashed-dotted line, showing that it converges with the full ratio (solid line) when the
distance $z_0$ is well above the screening length, i. e., $z_0{>\atop\sim}\sqrt{r_s\,a_0}$.

\section{Conclusions}
\label{sec:level5}

First of all, we have presented semi-classical and quantized hydrodynamic models to obtain the
quadratic electronic response of a semi-infinite electron gas. Then, we have derived explicit
expressions for the dynamic image potential experienced by charged particles traveling parallel
to a jellium surface, up to third order in the projectile charge. In the case of a stationary
charged particle the total quadratic contribution to the image potential has been found to be, at
large distances from the surface, twice as large as the quadratic contribution coming from the
surface plasmon field. Near the surface, the total quadratic contribution to the image potential of
a charged particle with $v=0$ has been found to be larger than the quadratic contribution from the
surface channel by a factor of $\sim 10$. As the speed of the moving charged particle increases,
linear contributions to the image potential coming from the bulk channel have been found to
decrease, and quadratic contributions from coupling with the bulk plasmon field are also expected
to decrease with increasing velocity.

Though nonlinear corrections are found to be more important far inside the solid\cite{Pitarke2}
than outside, our results indicate that the nonlinear image potential is enhanced with respect
to the linear image potential by a factor that is for aluminum as large as $\sim 1.15$ near the
surface in the case of a stationary charged particle ($v=0$) with unit charge $e$ ($Z_1=1$). At
large distances ($z_0>>\sqrt{r_s\,a_0}$) from the surface, the ratio between quadratic and linear
contributions to the image potential of a stationary charged particle decreases with the
distance $z_0$ as $Z_1\,1.93\times10^{-2}\,(z_0/r_s)^{-3}$, showing that it vanishes at high
electron densities.

As the speed of the moving charged particle increases, quadratic contributions to the image
potential are found to be very small. In particular, in the case of a projectile of charge
$Z_1=10$ moving with speed $v=2$ near the metal surface, contributions to the quadratic image
potential from coupling with the surface plasmon field have been found to enhance the linear image
potential near the surface by a a factor of $\sim 1.14$.

\acknowledgments
 A.B. and J.M.P. wish to acknowledge partial support by the basque Unibertsitate eta Ikerketa
Saila and the spanish Ministerio the Educaci\'on y Cultura.

\appendix
\section{}

Here we use both semi-classical and quantized hydrodynamic models to evaluate the first order
electron density $n_1$, which we assume to be induced by a stationary particle (${\bf v}=0$) of
charge $Z_1$ that is located far from the surface, i. e., $z_0\to\infty$.

Within the semi-classical hydrodynamic model, this quantity is easily found from Eq. (\ref{1.A.8})
to be given by the following expression
\begin{equation}
n_1(z)=Z_1\,{\omega_{\rm p}\over 2\pi\beta}\,{{\rm e}^{{\omega_{\rm p}
z}/\beta}\over z_0^2}\,\Theta(-z).\label{AP.1}
\end{equation}

Within the quantized hydrodynamic model, the operators corresponding to the induced electron
density due to bulk/surface plasmons are obtained as follows
\begin{equation}
\hat n^B({\bf r},t)=-{1\over\Omega}\sum_{{\bf q},p>0}g_{{\bf q},p}^B(z)
\,{\rm e}^{{\rm i}{\bf q}\cdot{\bf r}_\parallel}
\,\chi^B_{{\bf q},p}(t)\,\Theta(-z) \label{AP.2}
\end{equation}
and
\begin{equation}
\hat n^S({\bf r},t)=-{1\over A}\sum_{\bf q}\ g_{\bf q}^S(z)
\,{\rm e}^{{\rm i}{\bf q}\cdot{\bf r}_\parallel}
\,\chi^S_{{\bf q}}(t)\,\Theta(-z),\label{AP.3}
\end{equation}
the operators $\chi^B_{{\bf q},p}(t)$ and $\chi^S_{{\bf q}}(t)$ being given by Eqs. (\ref{2.14.A})
and (\ref{2.14.B}), respectively, and $g_{{\bf q},p}^B(z)$ and
$g_{\bf q}^S(z)$ representing bulk and surface coupling functions,
\begin{eqnarray}
g_{{\bf q},p}^B(z)=&&\left\{{2\,\omega_{\rm p}/\left(\pi\,\omega^B_{{\bf q},p}\right)\over
\left[2\,(\omega^B_{{\bf q},p})^2-\omega_{\rm
p}^2\right]^2-4\,\beta^2\,q^2\,(\omega^B_{q,p})^2}\right\}^{1/2}\cr\cr\cr
&&\times\left\{p\left[\omega_{\rm
p}^2+2\,\beta^2\,\left(p^2+q^2\right)\right]\,\cos pz\right.\cr\cr
&&\left.+q\,\omega_{\rm
p}^2\,\sin pz\right\}\label{AP.4}
\end{eqnarray}
and 
\begin{equation}
g_{\bf q}^S(z)=\sqrt{q\,\gamma_{\bf q}/(\omega^S_{\bf q}\,\pi)}\,\omega_{\rm p}\,
{q+\gamma_{\bf q}\over\left(q+2\,\gamma_{\bf q}\right)^{1/2}}\,{\rm e}^{\gamma_{\bf
q}\,z}.\label{AP.5}
\end{equation}

The electron density induced by the presence of the external perturbing charge is obtained as the
expectation value of the total electron density operator. Up to first order in the external
perturbation and in the limit as $z_0\to\infty$, we find
\begin{equation}
n_1(z)=n_1^B(z)+n_1^S(z)\label{eqnew},
\end{equation}
where
\begin{equation}
n^B_1(z)=Z_1\,{\omega_{\rm p}\over 2\pi\beta}
{{\rm e}^{\omega_{\rm p}\,z/\beta}-{\rm e}^{\omega_{\rm p}\,z/\left(\sqrt 2\,\beta\right)}/\sqrt
2\over z_0^2}\label{AP.6}
\end{equation}
and
\begin{equation}
n^S_1(z)=Z_1\,{\omega_{\rm p}\over 2\pi\beta}
{{\rm e}^{\omega_{\rm p}\,z/\left(\sqrt 2\,\beta\right)}/\sqrt
2\over z_0^2}.\label{AP.7}
\end{equation}
Eq. (\ref{eqnew}) coincides with Eq. (\ref{AP.1}), and shows that as long as the stationary
charged particle is located far from the surface the total electron density $n_1$ induced at the
surface ($z\sim 0$) is, in the limit as $z_0\to\infty$, $\sqrt 2$ times the electron density
$n_1^S$ induced through coupling of the charged particle with the surface plasmon field.

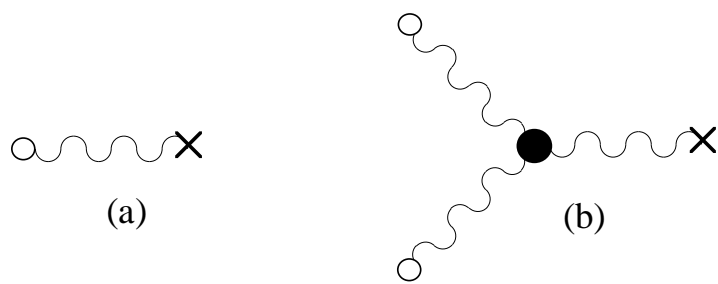
\begin{figure}
\caption{Feynman diagrams representing first (a) and second (b) order contributions to the electric
potential induced by an external charged particle. The external perturbation is represented by
white points, and the cross represents a test positive unit charge. Wavy
lines represent plasmon propagators and the black point, joining three plasmon lines,
describes the nonlinear interaction between three excitations}
\label{1}
\end{figure}

\begin{figure}
\caption{Linear contribution to the image potential of a particle with charge $Z_1=1$ and speed
$v=0$ (a) and $v=2$ (b) traveling parallel to the surface of a semi-infinite electron gas with
$r_s=2$, as a function of the distance $z_0$ from the surface. Solid lines represent the
full linear contribution to the image potential. Dashed and dotted lines represent
contributions from the excitation of bulk and surface plasmons, respectively. The classical image
potential of Eq. (\ref{1.A.16}) is represented by a dashed-dotted line.}
\label{2}
\end{figure}

\begin{figure}
\caption{As in Fig. 2, for the quadratic contribution to the image potential of a particle with
charge $Z_1=1$. Dotted lines represent the quadratic contribution from to
the image potential coming from surface plasmon excitation. In the case of a stationary particle
($v=0$), the full quadratic contribution to the image potential is represented by a solid line.}
\end{figure}

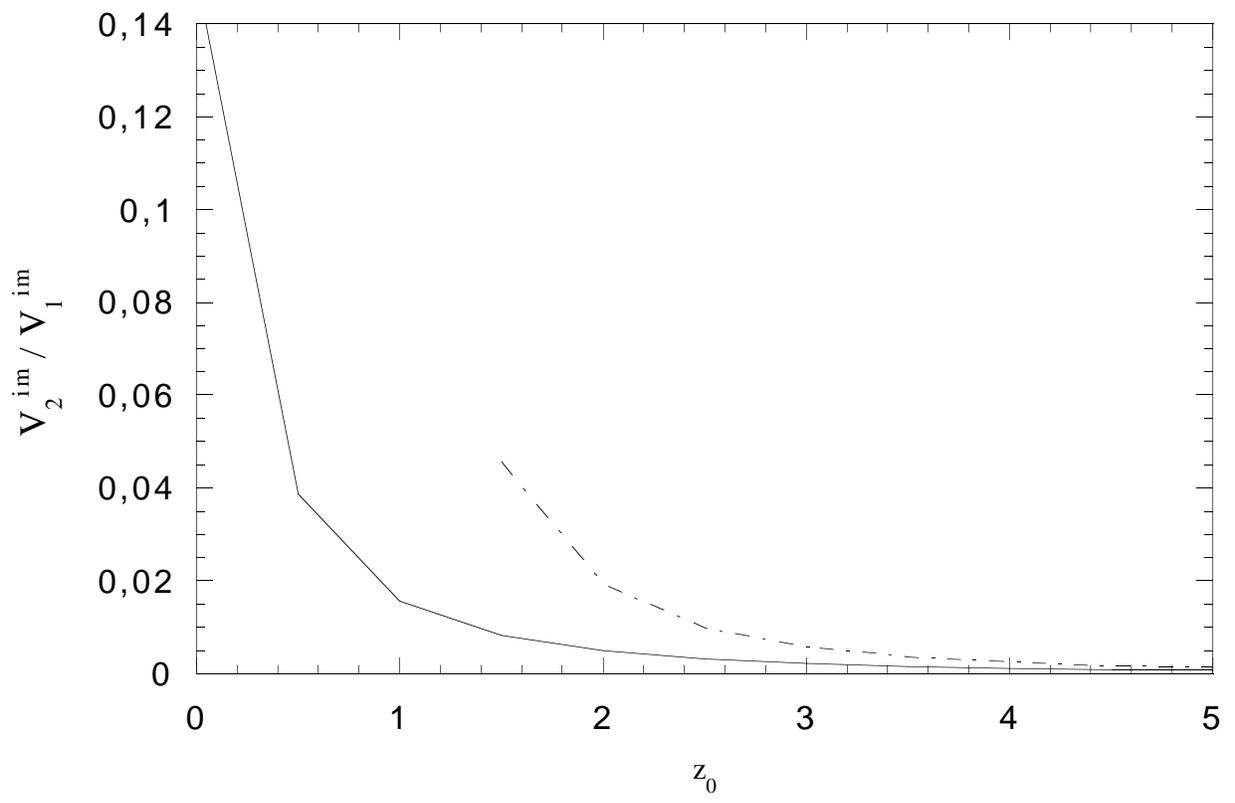
\begin{figure}
\caption{Ratio between quadratic and linear contributions to the
image potential of a stationary particle ($v=0$) with charge $Z_1=1$ that is located outside the
surface of a semi-infinite electron gas with $r_s=2$ at a distance $z_0$ from the surface. The solid
line represents the ratio between full calculations of quadratic and linear image potentials. The
dashed-dotted line represents the approximate ratio $Z_1\,1.93\times10^{-2}\,(z_0/r_s)^{-3}$, taken
from Eq. (\ref{1.B.18}).}
\end{figure}


\begin{references}
\bibitem{note1} The one-electron radius $r_s$ is defined by the relation $n_0=3/(4\,\pi\,r_s^3)$.
\end{references}
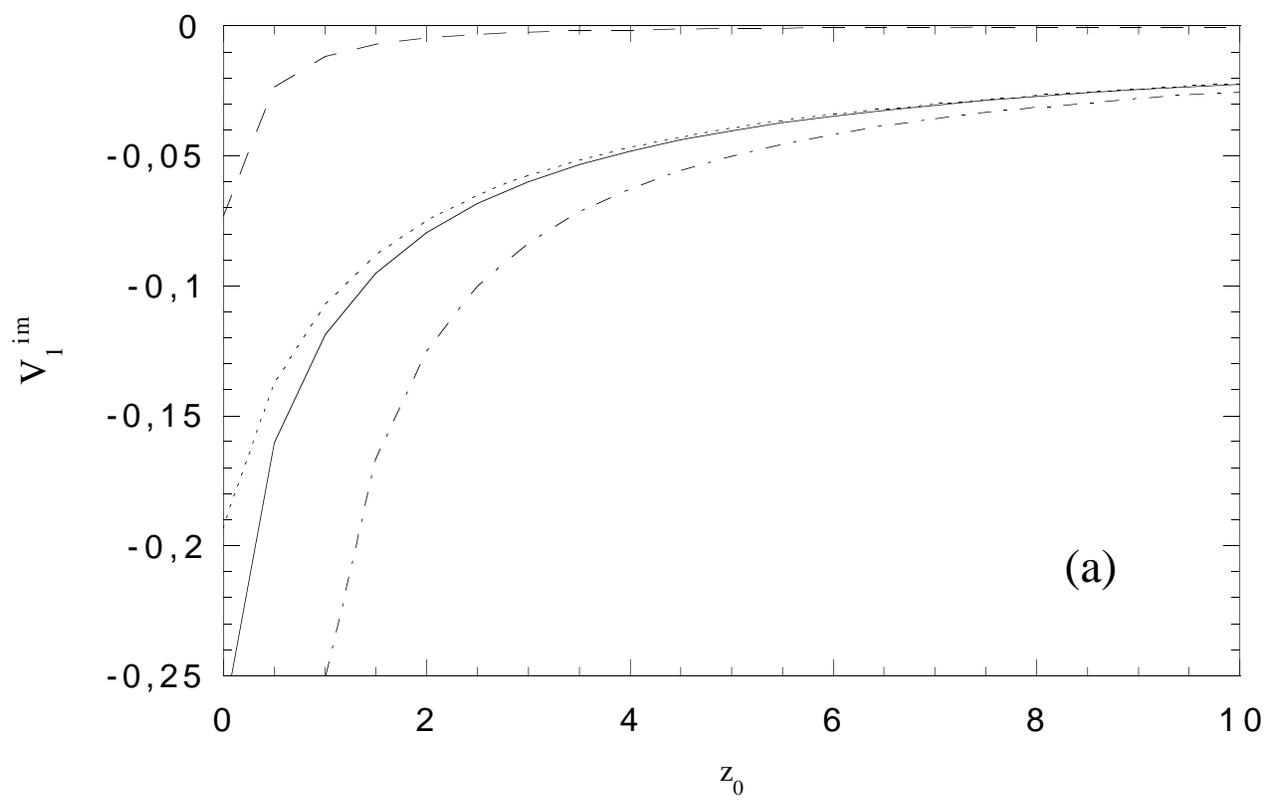
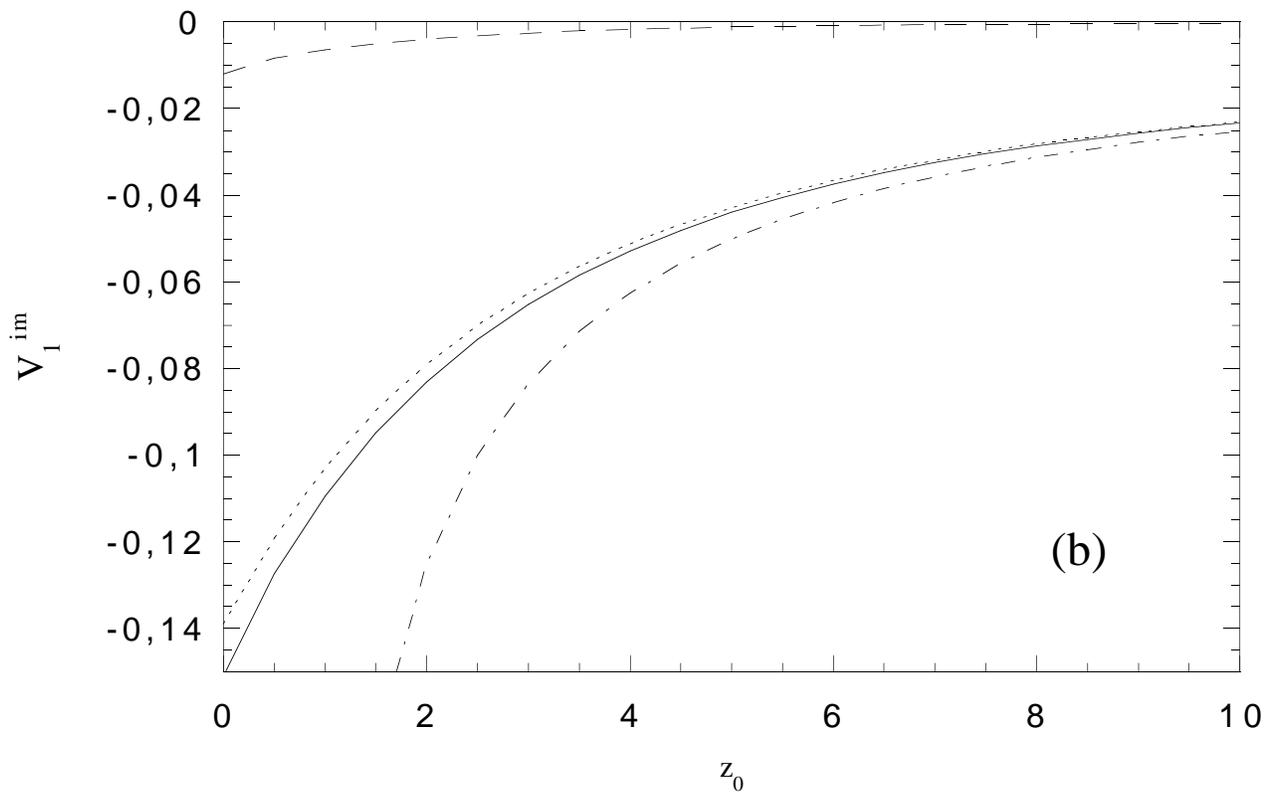
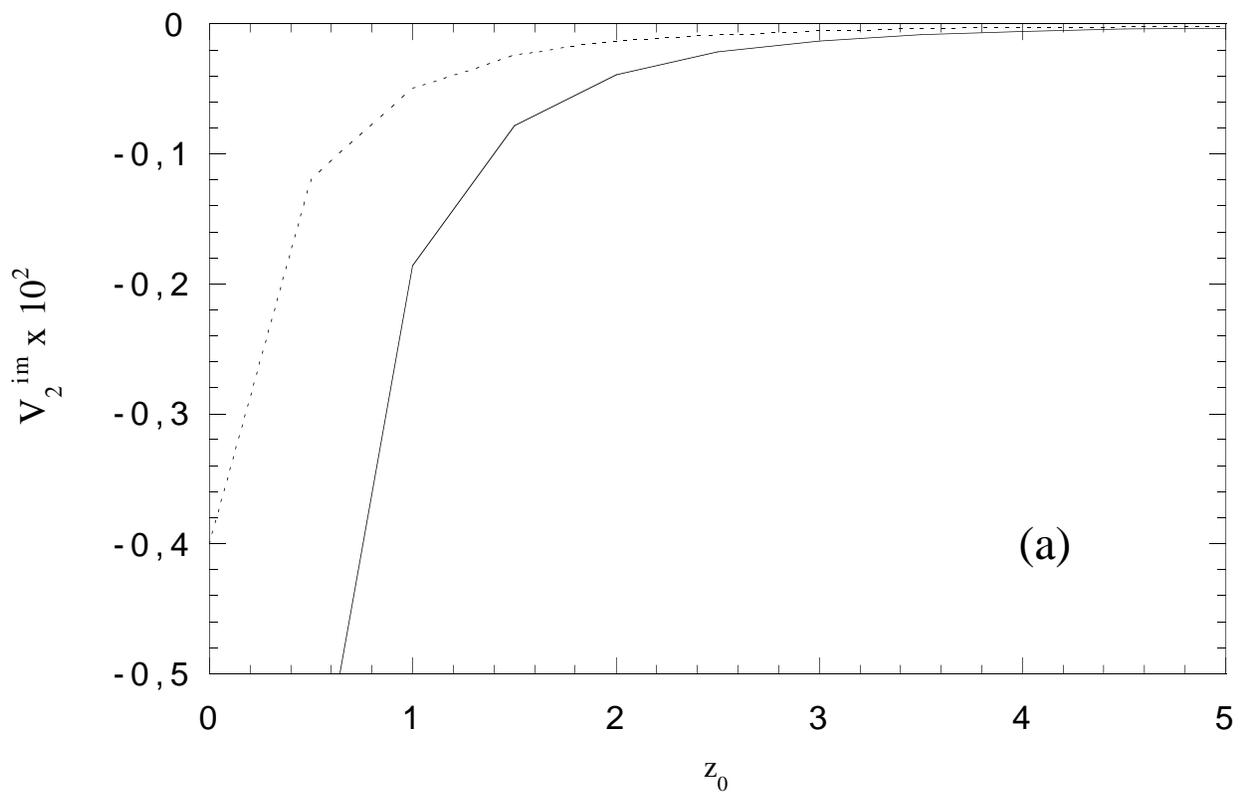
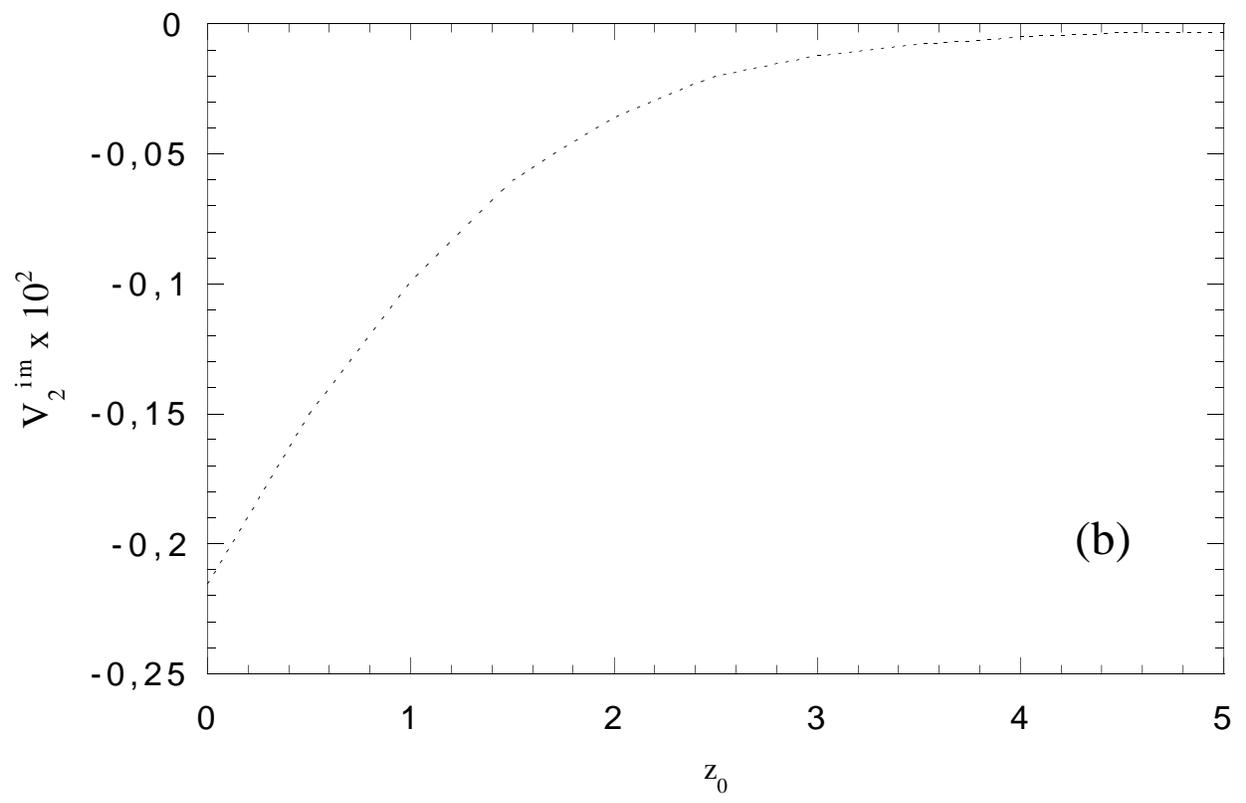
\end{document}